\documentstyle[aps,prb,12pt]{revtex} 
\title{Strongly localized anharmonic modes in perfect and imperfect 
crystals} 
\author{V.Hizhnyakov$^{1,3}$, D.Nevedrov$^{1,2}$, E.Sigmund$^3$} 
\address{$^1$Institute of Theoretical Physics, 
University of Tartu, T\"{a}he 4, 
Tartu EE2400, Estonia.\\
Institute of Physics, University of Tartu, Riia 142, Tartu EE2400, Estonia.\\ 
$^2$NORDITA, Blegdamsvej 17, Copenhagen DK-2100, Denmark.\\
$^3$Brandenburgische Technische Universit\"at Cottbus, Karl-Marx-Str. 17, 
D-03044 Cottbus, Germany.} 
\begin{document} 
\maketitle
\begin{abstract} 
Localized modes of large amplitudes in nonlinear lattices are considered. 
The applied method allows the reduction of nonlinear problem to a linear 
inverse problem of phonons scattering on a local potential. 
The method is efficient in the case of strongly localized modes. 
Analytical description of such modes in monatomic chain is given. 
Results of numerical calculations of anharmonic local vibrations of light ions
in pure and impure alkali halide crystal are presented. It is found that, 
in the case of amplitudes $\tilde{>} 0.5\AA$, the vibrations depend very 
strongly on the crystallographic directions.
\end{abstract} 



\section{Introduction} 
Recent studies of nonlinear vibrational dynamics lead to observation of 
long-living localized vibrations in perfect anharmonic lattices
\cite{ovchi,kosevich,dolgov,sivtak,page,zavt} (see also review \cite{flach} 
and references therein). The methods, which have been used so far for studying
of these novel excitations, are based on direct numerical integration of 
classical equations of motions, being here the nonlinear differential 
equations. Due to the fast growth of the number of numerical operations and 
computational
time with increase of the number degrees of freedom, the research is mainly 
focused on simple models, such as one-dimensional one- and two-atom  
lattices; only little have been done for real 3D crystals \cite{flach}. 
On the contrary, in harmonic approximation calculations can be performed also 
for macroscopically large pure and locally distorted crystals.  
Here we present a method which makes use of harmonic approximation results
for calculations of the anharmonic local modes (ALM). The developed 
techniques allow one to reduce the nonlinear problem of ALMs to a properly 
formulated problem of linear local dynamics. It is applicable for lattices of 
arbitrary dimension, being especially efficient in the case of strongly 
localized modes, when it allows to obtain analytical solutions. The 
consideration is based on classic mechanics; account of quantum effects is 
given in \cite{ovchi,hizhrev,hizhnev,zhizh,chain,euro}. 

The starting points for the method are the stability conditions of an
ALM with respect to small fluctuations:
\begin{itemize}
\item
all frequencies of the spectrum of these fluctuations should be real and  
positive,  
\item
spectrum of fluctuations should contain also the basic frequency $\omega_l$ of
the ALM, i.e. there should be harmonic local mode (HLM) with the same 
frequency $\omega_l$; the same holds for the higher harmonics of the mode, 
\item
the ratios of the amplitudes of atoms of the HLM should coincide with that of 
corresponding ALM.  
\end{itemize}

The physical meaning of the condition a) is obvious: the configuration of the 
crystal lattice with the ALM must be stable (not metastable). 
The conditions b) and c) ensure that a solution, which describes  
an ALM are stable  with respect to infinitesimal changes of the ALM phase. 
The last two conditions allow the self-consistent calculation of the ALMs.
Indeed local mode of finite amplitude in anharmonic lattice
causes local perturbation of the dynamical matrix of small vibrations. 
Therefore, knowing (or choosing as a trial parameters) the amplitudes of 
the ALM on different atoms, we can determine the changes of the 
elastic springs, caused by the ALM. Then, by applying  the Lifschitz formula
of local dynamics in harmonic approximation, the expressions for the
HLM (and ALM) may be obtained, both for its frequency and amplitudes of 
contributing atoms. These expressions allow the obtaining
of the self-consistency conditions, being the set of of equations for the 
parameters of the ALM. Thus, the method  reduces the nonlinear problem of 
calculation of an ALM to a kind of a linear inverse problem of phonons 
scattering on a local scattering potential for small vibrations (phonons). 
If the ALM is strongly localized, then 
the procedure is simple while one needs to calculate only few amplitudes. 
This can be done analytically. The method is applicable for lattices of 
different dimensions. Besides, it can be also used for calculations of 
anharmonic local modes of defects. 

To illustrate, how the method works, we will apply it first to the ALMs in 
monatomic chain 
with hard quartic anharmonicity. We will show that one gets analytical results
which, in the limiting case of high frequency, coincide with known  
solutions \cite{page}. Then we will calculate odd ALMs in pure and impure  
alkali halide crystals with light and heavy ions and with account of   
quartic anharmonicity (the ALMs under consideration are almost fully
localized on the light atoms).  Frequency spectra in these crystals are 
calculated in the shell model \cite{bilz,shell,maradu,kristofel}.
Anharmonic constants are determined from the Born-Coulomb-Mayer potentials. 

\section{General} 
Let us consider a vibrational system with the potential energy  
\begin{equation} 
{V} = \frac{1}{2} \sum_{n_{1} n_{2}}\!{V^{(2)}_{n_{1} n_{2}} 
{U}_{n_{1}} {U} _{n_{2}}} 
+ \frac{1}{3}\! \sum_{n_{1} n_{2} n_{3}}\!\!{V^{(3)}_{n_{1} n_{2} n_{3}} 
{U}_{n_{1}} {U}_{n_{2}} {U}_{n_{3}}} + \ldots\,, 
\label{eq:v} 
\end{equation} 
where ${U}_{n}$ are the Cartesian displacements of atoms situated at the site 
$n$, $V^{(2)},\, V^{(3)},\ldots$ are harmonic and anharmonic springs. The 
displacements ${U}_{n}$ satisfy the following equations of motion 
\begin{equation}
-M_{n} \ddot{U}_{n} =  
\sum_{n_{1}}{V^{(2)}_{n n_{1}} {U}_{n_{1}}} +  
\sum_{n_{1} n_{2}} V^{(3)}_{n_{1} n_{2} n_{3}} {U}_{n_{1}} 
{U}_{n_{2}} + \ldots,
\end{equation}
where $M_{n}$ are the masses of atoms. We suppose that a ALM is excited
at the site $n=0$ and its nearest neighbors. This excitation is the solution 
of the equations of motion with exponentially decreasing  $|U_n|$ 
with $|n| \rightarrow \infty$. As it is known \cite{dolgov,sivtak,flach} a
solution, which corresponds to the ALM with the frequency  $\omega_l$ has the 
form $U_n(t)= A_n \cos(\omega_l t) + \xi_n + O(\omega_l)$, where $O(\omega_l)$ 
is the sum of small terms of the frequencies $3\omega_l$,  
$5\omega_l, \ldots$ (i.e. the higher order harmonics; these harmonics are 
neglected below).  The amplitudes $A_{n}$ and the shifts $\xi_{n}$ can be 
found by equating the coefficients before the terms with the same time 
dependence \cite{dolgov,sivtak,flach}. The corresponding equations are
\begin{eqnarray} 
M_{n} \omega_{l}^{2} U_{n} &=& \sum_{n1}{V^{(2)}_{n n_{1}} U_{n_{1}}} 
+ 2 \sum_{n_{1} n_{2}}{V^{(3)}_{n n_{1} n_{2}} U_{n_{1}} \xi_{n_{2}}} 
\nonumber \\ 
&+& 3\sum_{n n_{1} n_{2} n_{3}}{V^{(4)}_{n n_{1} n_{2} n_{3}} 
{\Big (}\frac{1}{4}U_{n_{1}} U_{n_{2}} + \xi_{n_{1}} \xi_{n_{2}}{\Big )} 
U_{n_{3}}} + \ldots, 
\label{eq:a}
\end{eqnarray} 
$$
\sum_{n_{1}}{V^{(2)}_{n n_{1}} \xi_{n_{1}}} + 
\sum_{n_{1} n_{2}}{V^{(3)}_{n n_{1} n_{2}} {\Big (}\frac{3}{2} U_{n_{1}} 
U_{n_{2}} + \xi_{n_{1}} 
\xi_{n_{2}}{\Big )}} + \ldots = 0.
$$
To take into account small fluctuations we present the displacements in the  
form $\bar{U}_n = U_n + q_n/ \sqrt{M_n}$, where $q_n $ stands for the reduced  
small displacement. The equations for ${q}_{n}$ are obtained from
(\ref{eq:a}) and read: 
\begin{eqnarray}
\ddot{q}_{n}=  
\sum_{n'}{{\big (}V_{2 n n'}+W_{2 n n'}{\big )}{q}_{n'}},
\label{eq:b}
\end{eqnarray}
where $V_{2 n n'}$ is the dynamical matrix in harmonic approximation, 
\begin{eqnarray} 
W_{2 n n'} & = & \frac{2}{\sqrt{M_{n} M_{n'}}}{\Big (} 
\sum_{n'}{V^{(3)}_{n n' n_1}\xi_{n_1}} + \nonumber \\ 
& + & \frac{3}{4} \sum_{n_{1} n_{2}}{V^{(4)}_{n n'n_{1} n_{2}}{\big 
(}U_{n_{1}} U_{n_{2}} + 2\xi_{n_{1}} \xi_{n_{2}}{\big )}}  
+ \ldots\ {\Big )} ,
\label{eq:c}  
\end{eqnarray}  
describes anharmonic renormalization of the dynamic matrix due to the ALM 
(time-dependent terms $\sim \cos{k \omega_{l} t}$, $k=0,1,2,\ldots$ are 
neglected here; these terms are important in quantum mechanical description
of fluctuations: they lead to multi-phonon decay of the ALM
\cite{hizhrev,hizhnev}).
In the case of strongly localized modes one can include into consideration
only small number $N$ of amplitudes $U_n$. Then $\{W_{2nn'}\}$ is the 
$N$x$N$-matrix of the local springs of the lattice.

Hamiltonian, which corresponds to the 
equations of motions (\ref{eq:b}), equals 
$H_{ph} = H_{0,ph} + \tilde{V}_2,$ where 
\begin{equation}
H_{0,ph} = \frac{1}{2} \sum_n \dot{q}_n ^2 + \frac{1}{2} \sum_{nn'}   
V_{2nn'} q_n q_{n'}= 
\frac{1}{2} \sum_{i}(\dot{x}_i^2+\omega_i^2 x_i^2) 
\end{equation}
is the phonon Hamiltonian of the harmonic lattice, 
$x_i = \sum_n e_{in}q_n$ are the normal coordinates of the lattice 
(in a perfect lattice $e_{in} \sim e^{i\vec{k} \vec{n}}$), 
$$
\tilde{V}_2  = \frac{1}{2} \sum_{nn'}W_{2 n n'}q_n q_{n'} 
$$
describes the effect of the ALM on the phonons, 
$W_{2 n n'}$ are corresponding local distortions of the  
elastic springs, given by (\ref{eq:c}).  

The HLMs, induced by the interaction $\tilde{V}_2$, are 
determined by the poles of the spectral Greens functions 
$$ 
G_{nn'}(\omega) = -i \int_{0}^{\infty} e^{i\omega t - \epsilon t} 
G_{nn'}(t)
$$
on the real axis of $\omega$;
$G_{nn'}(t) = \Theta(t) \sum_i \bar{e}_{in} \bar{e}_{in'} \sin(\omega_i t)$  
is the retarded matrix-Green's function of phonons,
$\bar{e}_{in} = e_{in}/\omega_i$ . 
This  matrix-function can be found by applying Lifshitz formula 
\begin{equation}
G_{nn'} = G_{nn'}^{(0)} + \sum_{n_1 n_2} G^{(0)}_{nn_1}W_{2n_1 n_2} 
G_{n_{2} n' },
\end{equation}
where $G_{nn'}^{(0)}(\omega)$ are the Green's functions of the harmonic
lattice. These functions can be calculated by standard methods of lattice 
dynamics. To find the frequency and shape of the HLM (and ALM) let us 
introduce the configurational coordinates 
$$
Q_{\nu} = \sum_{n} S_{\nu n} q_n, \hspace{5mm}
\tilde{Q}_{\mu} = \sum_{\nu} s_{\mu \nu} (\omega) Q_{\nu}. 
$$
The first 
transformation $S$ is chosen to diagonalize the perturbational quadratic form 
$\tilde{V}_2$: 
$$
\tilde{V}_2 = \sum_{\nu} \eta_{\nu} Q_{\nu}^2 /2.
$$
The second transformation $s(\omega)$ diagonalizes the dimensionless 
Green's function matrix $\check{G}_{\nu \nu'}=
\sqrt{\eta_{\nu} \eta_{\nu'}}\bar{G}^{(0)}_{\nu\nu'}$: 
$$
\tilde{G}^{(0)}_{\mu \mu'} =  \delta _{\mu \mu'} 
 \sum_{\nu \nu'} s_{\mu \nu} s_{\mu \nu'} \check{G}_{\nu \nu'} \, =
\, \delta _{\mu \mu'} \sum_{nn'} R_{\mu n}R_{\mu n'} G^{(0)}_{n n'}.
$$ 
Here  $\bar{G}^{(0)}_{\nu\nu'}(\omega)$ are the lattice Greens functions  
in $Q_{\nu}$-space, $R_{\mu n} (\omega) =  
\sum_{\nu} s_{\mu \nu}(\omega) \sqrt{\eta_{\nu}} S_{\nu n}$. 
In the $\tilde{Q}_{\mu}$-representation the Lifshitz formula takes the form 
$\tilde{G}_{\mu \mu}(\omega) =  
\tilde{G}^{(0)}_{\mu \mu}(\omega)(1 -\tilde{G}_{\mu \mu}(\omega)).$

From symmetry considerations one can chose the configurational 
coordinate $\tilde{Q}_{\mu}$ which corresponds to the ALM under investigation. 
The frequency $\omega_l$ of the local mode is given by the 
position of the pole of the Green's function $\tilde{G}_{\mu \mu}(\omega)$
on the real axis $\omega$, i.e. satisfies the relation
\begin{equation} 
\tilde{G}^{(0)}_{\mu \mu}(\omega_l) = 1; 
\end{equation} 
the relative amplitude of the $n$' atom in the mode is given by the formula: 
$a_{\mu n} = R_{\mu n}(\omega_l) \sqrt{\pi / |G^{(0)'}_{\mu \mu} (\omega_l)|},$
where  
$G^{(0)'}_{\mu \mu}(\omega) = d G^{(0)}_{\mu \mu}(\omega)/ d \omega$. 
Taking into account that the relative amplitudes satisfy the 
normalization condition  $\sum_n a_{\mu n}^2 = 1$, we obtain: 
\begin{equation} 
a_{\mu n} = 
\frac{R_{\mu n}(\omega_l)}{\sqrt{|\sum_{\nu} s^2_{\mu \nu}(\omega_l)
\eta_{\nu}|}}. 
\label{eq:e}
\end{equation} 
The relations (\ref{eq:e}) give the set of equations for $a_{\mu n}$. 
In combination  with (8)
they allow one to find the amplitudes $A_{n}$ of the ALM of 
given frequency $\omega_l$. The method works better for strongly localized 
ALMs with high frequency, when the number of contributing amplitudes $A_n$
(and equations to be solved) is small. 

In the case of high frequency of the ALM  ($\omega_l \gg \omega_M$, $\omega_M$ 
is maximal phonon frequency) the non-diagonal elements of the  
$G^{(0)}_{nn'}(\omega_l)$ are much smaller than the diagonal ones. Then 
$\check{G}^{(0)}_{\nu \nu'}(\omega_l) \approx \delta_{\nu \nu'}  
G^{(0)}_{00}(\omega_l)$, i.e. the Greens-function matrix 
diagonalizes simultaneously with $\tilde{V}_2$. In this case the normalized
amplitudes of the ALM can be determined by simple self-consistency 
condition $a_n =  S_{\nu n}$; absolute amplitudes one can  find from the
relation $G_{\nu \nu}(\omega_l) = \eta_{\nu}^{-1}$. 

If the limited number of atomic displacements is taken into consideration
then the quadratic form
$\tilde{V}_2$ does not  depend on the totally-symmetric coordinate, being the 
common displacement (the sum of equal displacements of all accounted atoms). 
This means that all modes, which contribute to the diagonalized quadratic form
$\tilde{V}_2$, are orthogonal to this displacement. Evidently, the same holds 
for the liner combination of shifts, which give contributions to the ALM. 
Otherwise, the sum of displacements of all atoms, which contribute to the ALM,
should be equal zero: 
\begin{equation}
\sum_n A_n = 0. 
\end{equation}
The same holds for the sum of 
momenta of atoms at any time-moment. This property of the ALMs may be called 
as immobility condition.  

\section{Even and odd ALMs in monatomic chain} 
We apply the method to description of ALMs in monatomic chain. The potential 
energy of the chain with account of the nearest-neighbor interactions has the 
form 
$$ 
V=\sum_n \sum_r \frac{V^{(r)}}{r} (U_{n+1} -U_n)^r. 
$$ 

In harmonic approximation ($V^{(2)}>0,\,\, V^{(r)}=0,\,\, r\geq 3$) for 
$\omega/\omega_M > 1$ the Green's functions of the chain equal 
\cite{economou}
$$
G^{(0)}_{nn'}(\omega) = \frac{(-\rho)^{|n-n'|}\omega_M/\omega}
{4\omega_M \sqrt{\omega^2 - \omega_M^2}},
$$ 
$\omega_M = 2\sqrt{V^{(2)}/M_0}$ is the top phonon frequency,
$\rho = (\omega/\omega_M - \sqrt{\omega^2/\omega_M^2-1})^2 < 1$.
In the limit of strong amplitudes and high frequency one gets two strongly 
localized  ALMs \cite{sivtak,page}: 
\begin{enumerate}
\item the odd mode with $A_n= A_{-n}$,  
$A_1 \simeq -A_0/2$, $|A_n|\ll |A_1|$, $n \ge 1$ and 
\item the even mode with 
$A_{n+1} = -A_{-n}$, $|A_{n+1}| \ll |A_0|$, $n \ge 1$. 
\end{enumerate}
For $n\gg 1$ we have $A_{|n|+1} \simeq -\rho A_{|n|}$.
For smaller frequency the wings (tails) of the mode
are stronger. Below we give analytical description of strongly localized ALMs,
having remarkable $\omega_l / \omega_M -1,$ in the chain with  quartic  
anharmonicity ($r=4$). Note that in the opposite limit of very small   
$\omega_l/\omega_M -1$ there exists an analytical solution of the problem,
which describes the ALM of large size \cite{kosevich}.

\subsubsection{Even mode}
Let us consider first the even mode. With account of four central atoms 
displacements the contribution of the ALM to the potential energy of phonons 
has the form 
$$
\tilde{V}_2 = K_4 A_0^2 [b((q_2-q_1)^2+(q_0 -q_{-1})^2)+(q_1 - q_0)^2], 
$$
where $K_4= 6V^{(4)}/M_0$, $b=(1+\beta)^2/4$, $\beta = A_2/A_0$.  
Two following even modes contribute to the $\tilde{V}_2 $ and ALM: 
$$
y_1 = {q_0-q_1}{\sqrt{2}},\quad y_2 = {q_2-q_{-1}}{\sqrt{2}}.
$$
In the strong localization limit the main contribution is given by $y_1$. 
Contribution of $y_2$ depends on the $\omega_l$ being larger for smaller 
$\omega_l$. Our task is to describe the last dependence. In $y_1,\,y_2-$ 
subspace  $\tilde{V}_2 = K_4((2+b)y_1^2 - 2b y_1y_2 + b y_2^2)$. 
This quadratic form is diagonalized in the rotated basis 
$Q_1 = y_1\cos{\phi} +  y_2\sin{\phi}, \quad 
Q_2 = - y_1\sin{\phi} +  y_2\cos{\phi}$
with $\tan{(2\phi)} =b$.
In the limit of high frequency $\omega_l$ the self-consistency condition  
reads 
$$
\sin{\phi}=\beta \approx \frac{b}{2} = \frac{(1+\beta)^2}{8}.
$$
One gets 
$\beta \approx 1/6$ in agreement with the corresponding calculation of $\beta$
on the basis of the nonlinear equations of motion \cite{page}. 
To describe $\omega_l-$dependence of $\beta$ we should find coordinates 
$\tilde{Q}_{1,2}$ which diagonalize the Green's function matrix 
$\check{G}^{(0)}_{\nu \nu'}= 
G^{(0)}_{00}(\sqrt{\eta_{\nu}\eta_{\nu'}}g_{\nu\nu'})$, $\nu, \nu' =1,2$,
$\eta_{1,2} =K_4 A_0^2 (1+b \pm \sqrt{1-b^2})$,
\begin{eqnarray}
g_{11} &=&\rho_{11}\cos^2{\phi} +
\rho_{22}\sin^2{\phi} + \rho_{12} \sin{(2\phi)}, \nonumber \\
g_{22} &=&\rho_{11}\sin^2{\phi} +
\rho_{22}\cos^2{\phi} - \rho_{12} \sin{(2\phi)}, \nonumber \\
g_{12} &=&-\rho_{12} \cos{(2\phi)} + \frac{1}{2}(\rho_{11}- \rho_{22})
\sin{(2\phi)},  \nonumber
\end{eqnarray}
where
$\rho_{11} = 1+\rho$, $\rho_{22} = 1+\rho^3$, $\rho_{12}=\rho(1+\rho)$.
The diagonalization is achieved by rotating of the coordinates $Q_{1,2}$ on 
the angle $\alpha$ with 
\begin{equation}
\tan{(2\alpha)}= \frac{2\sqrt{\eta_1 \eta_2}g_{12}}
{(\eta_1 g_{11}-\eta_2 g_{22})}.
\label{eq:f}
\end{equation}
The self-consistency condition now is 
\begin{equation}
\beta = \sin{(\alpha +\phi)}.
\end{equation}
This is equation for $\beta(\omega_l)$, which can be easily solved numerically
or (approximately) analytically. E.g.
in the strong localization limit, when both, $\alpha$ and $\phi$ are small,
one gets 
\begin{equation}
\beta \approx  \frac{1}{6} + \frac{5\rho}{12\sqrt{2}(1+\rho)}. 
\end{equation}

\noindent
The dependence of the dimensionless frequency of the mode 
$\Omega_l = \omega_l/\omega_M$ on the amplitude $A_l$ is given by the equation
(8) with $\nu =1$:
\begin{equation} 
\tilde{G}^{(0)}_{11}(\omega_l)= K_4 A_0^2G^{(0)}_{00}(\omega_l) 
{\Big (}1+b + \sqrt{1-b^2}{\Big )}
{\Big [}g_{11}\cos^2{\alpha} + g_{22} \frac{\eta_2}{\eta_1}\sin^2{\alpha} + 
g_{12}\sqrt{\frac{\eta_2}{\eta_1}}\sin{(2\alpha)}{\Big ]} = 1. 
\end{equation}
For $\omega_l > 1.15 \omega_M$ given formulas describe the ALM rather well:
the contributing amplitudes of the next nearest atoms is less 
than $10^{-2}$, thereat their contribution to the energy of the ALM is less 
than $10^{-4}$. 

\subsubsection{Odd mode}
Let us consider now the odd ALM with account of displacements of five atoms. 
The mode under consideration satisfies the parity and immobility conditions 
$A_n=A_{-n}$, $A_2+A_1+A_0+A_{-1}+A_{-2}=0$, which give 
$A_1=-A_0 (1/2 + \beta)$, $\beta = A_2/A_0$. The contribution of the ALM to 
the potential energy of phonons equals 
$$ 
\tilde{V}_2 = \bar{K}_4 A_0^2 {\Big [}
\bar{b}((q_2-q_1)^2 + (q_{-2} - q_{-1})^2) +  
(q_1-q_0)^2 + (q_{-1} -q_0)^2{\Big ]}, 
$$ 
where $\bar{K}_4=9V^{(4)}(1+2\beta/3)^2/8$, 
$\bar{b}=(1+4\beta)^2/(3+2\beta)^2$. Two odd modes, which give contributions 
to the ALM read: 
$$ 
y_1=\frac{1}{\sqrt{6}}(2q_0 -q_1-q_{-1}),\,\,\,\, 
y_2=\frac{1}{\sqrt{30}}(3q_2 + 3q_{-2} - 2q_0 - 2q_1- 2q_{-1}). 
$$ 
In the $y_{1,2}$-space  $\tilde{V}_2 = \bar{K}_4 [(3+\bar{b}/3)y_1^2 -  
(2\sqrt{5}/3)\bar{b} y_1 y_2 + (5/3) \bar{b} y_2^2]$. This quadratic form is 
also diagonalized in the rotated basis with $\sin{(\phi)}\approx b/2$,
where $b = 2\sqrt{5}\bar{b}/(9-4\bar{b})$. In the limit
of large $\omega_l/\omega_M$ the main contribution to the ALM is given by
the $y_1$-mode; contribution of the $y_2$-mode is given by the self-consistency
condition 
$\beta \approx 3\sin{(\phi)}/ 2\sqrt{5} \approx \bar{b}/6$. Taking 
$\bar{b} \approx 1/9$ one gets the value $\beta = 1/54$ in agreement 
with \cite{page}. A more 
accurate approximation is $\beta \approx 3/131$. 

The $\omega_l-$dependence of 
$\beta$ and $A_0$- dependence of $\omega_l$ can be found in the same way 
as for the even mode. In this case 
$$
\eta_{1,2} = \frac{K_4A_0^2}{2\sqrt{5} + 4b} (3\sqrt{5} +15 b + 
(3\sqrt{5} + 9 b))(1+ \pm \sqrt{1-b^2}).
$$
For small $\beta$ one gets
$$
\eta_1 \approx K_4A_0^2{\Big (}\frac{8}{3} - \frac{40\beta}{18}{\Big )},
\quad \eta_2
\approx\eta_1 {\Big (}\frac{1}{16} + \frac{27\beta}{64}{\Big )}.
$$
The factors $g_{\nu \nu'}$ are also determined by the same expressions 
as in the even ALM case but
with the following $\rho_{\nu\nu'}$:
\begin{eqnarray}
\rho_{11} &=& 1+ \frac{4\rho}{3} +\frac{\rho^2}{3},\nonumber \\
\rho_{22} &=& 1+ \frac{4\rho}{15} - \frac{8\rho^2}{15} + 
\frac{4\rho^3}{5} + \frac{3\rho^4}{5},
\nonumber \\
\rho_{12} &=& \sqrt{5}\rho{\Big (}\frac{1}{3} + 
\frac{8\rho}{15} + \frac{3\rho^2}{15}{\Big )}. \nonumber
\end{eqnarray}
The self-consistency condition in this case is:
\begin{equation}
\beta =\frac{3 \sin{(\phi+\alpha)}}{2\sqrt{5} \cos{(\phi+\alpha)} 
-2\sin{(\phi+\alpha)}},
\end{equation}
where $\alpha$ is determined by (\ref{eq:f}); thereat  expressions for
$\eta_{1,2}$  and for $\rho_{\nu\nu'}$ are given above,
$g_{\nu\nu'}$ are also determined by the same formulae as in the even ALM case.
In the case $\beta \ll 1$ one gets 
$\sin{(\alpha)} \approx \tilde{\rho} \sqrt{\eta_2/5\eta_1}$
where $\eta_1 \approx 3-\bar{b}/3$, $\eta_2 \approx 5\bar{b}/3$, 
$\tilde{\rho}= \rho (5-3 \rho)/(3-\rho).$  This gives
\begin{equation} 
\beta \approx \frac{0.023 + \tilde{\rho}/9}{1-4 \tilde{\rho}/9}. 
\end{equation}
The dependence of the frequency of the mode  is given by  equation (6)
If $\omega_l > 1.15 \omega_M$ then the odd ALM is well localized 
($\beta < 0.065$),  more than 99.98 percent of energy of the mode 
come from 4 central atoms. 

The calculated dependences of $\Omega_l$ and $\beta$ on amplitude
for even and odd modes are plotted on Fig.1.

\section{ALMs in alkali halides} 
We present here calculations of odd local modes associated with light
impurity and host ions in alkali-halide crystals. In 
these cases the ALM is almost fully localized on the light ion and the 
problem reduces to calculation of the frequency of the mode in dependence on 
its amplitude.

Within the approximation of the nearest neighbors interaction the potential 
operator has the form 

\begin{equation}
\hat{V} = \sum_{\bar{\alpha}}\sum_{\vec{n}}\sum_{m=1}^{\infty} \frac{1}{2m} 
V^{(m)}_{\vec{n}_{\bar{\alpha}}} 
{\big (} \hat{R}_{\vec{n}_{\bar{\alpha}}} {\big )}^{m}\,,
\end{equation}
where $\bar{\alpha} = \pm x,\:\pm y,\:\pm z$ are the directions to the nearest 
neighbors, 
$\vec{n} = (n_{x},n_{y},n_{z})$ is the vector of the lattice sites, 
$\vec{n}_{\bar{\alpha}}$ is the vector of the site nearest to $\vec{n}$ 
in $\bar{\alpha}$ direction, 
$V^{(m)}_{\vec{n}_{\bar{\alpha}}} = V_m$ is the $m$-th derivative of 
the pair potential between atoms (ions) $\vec{n}$ and $\vec{n}_{\bar{\alpha}}$ 
at their mean distance $R_{0\vec{n}_{\bar{\alpha}}},\,\,\, 
\hat{R}_{\vec{n}_{\bar{\alpha}}} = [(R_{0 n_{\bar{\alpha}}} + \hat{r}_{\alpha 
\vec{n}_{\bar{\alpha}}})^{2} + \hat{r}^{2}_{\vec{n}_{\bar{\alpha}}} -
\hat{r}_{\alpha \vec{n}_{\bar{\alpha}}}^{2}]^{1/2} - 
R_{0\vec{n}_{\bar{\alpha}}}$ is the operator of distance between 
the nearest neighbors in the $\bar{\alpha}$-direction, 
$\hat{r}_{\beta\vec{n}_{\bar{\alpha}}}=q_{\beta\vec{n}}-q_{\beta 
\vec{n}_{\bar{\alpha}}}$, $q_{\beta}$ is the $\beta$-component of the 
displacement vector $\vec{q}_{\vec{n}}$ of the atom $\vec{n}$, $\,\,\alpha,
\beta = x,y,z,\,\,$ 
$\hat{r}_{\vec{n}_{\bar{\alpha}}}^{2} = \hat{r}^{2}_{x\vec{n}_{\bar{\alpha}}} +
\hat{r}^{2}_{y\vec{n}_{\bar{\alpha}}} + 
\hat{r}^{2}_{z \vec{n}_{\bar{\alpha}}}$.
By expanding $\hat{V}$ in the power series of displacement operators 
$\hat{r}_{\alpha\vec{n}_{\bar{\alpha}}}$one gets
\begin{eqnarray}
\hat{V} &=& \frac{1}{4}\sum_{\bar{\alpha} , \vec{n}}
[V_2\hat{r}^2_{\alpha \vec{n}_{\bar{\alpha}}} + 
V'_2(\hat{r}^2_{\vec{n}_{\bar{\alpha}}}-
\hat{r}^2_{\alpha \vec{n}_{\bar{\alpha}}}) +
\nonumber\\
&&\frac{1}{3}V_3\hat{r}^3_{\alpha n_{\bar{\alpha}}} + 
V'_3 \hat{r}_{\alpha n_{\bar{\alpha}}}(\hat{r}^2_{\vec{n}_{\bar{\alpha}}}-
\hat{r}^2_{\alpha\vec{n}_{\bar{\alpha}}})+
\frac{1}{12} V_4\hat{r}^4_{\alpha n_{\bar{\alpha}}} + \nonumber \\
&&\frac{1}{2}V'_4 \hat{r}^2_{\alpha n_{\bar{\alpha}}} 
(\hat{r}^2_{n_{\bar{\alpha}}}- \hat{r}^2_{\alpha n_{\bar{\alpha}}}) +
\frac{1}{4} V'' (\hat{r}^2_{n_{\bar{\alpha}}}-
\hat{r}^2_{\alpha n_{\bar{\alpha}}})^2 +...],
\end{eqnarray}
where
\begin{eqnarray}
V'_2 &=& V_1 R_0^{-1},\quad V'_3 = (V_2 - V'_2)R_0^{-1}, \nonumber \\
V'_4 &=& R_0^{-1}V_3 - 2R_0^{-2}(V_2 - V'_2),\quad
V''_4 = R_0^{-2} (V_2 - V'_2),
\end{eqnarray}
$V_{2}$, $V_{3}$ and $V_{4}$ make account of the central, while $V'_{2}$, 
$V'_{3}$, $V'_{4}$ and $V''_{4}$ of the non-central forces. The potential 
considered does not take account of the covalent interaction which leads to 
the chemical bonding. This (covalent) interaction can, however, be easily 
included in calculations by introducing additional terms of the type $V_{2}$, 
$V_{3}$ and $V_{4}$.

As it is known, harmonic non-central springs $V'_2$ are normally $5$ to $10$ 
times smaller than the central springs $V_2$. Our calculations of $V_3$ and 
$V'_3$ for alkali halides show that $V'_3/V_3$ is even smaller than $V'_2/V_2$ 
($V'_3$ is twenty-thirty times smaller than $V_3$). 
The same holds also for quartic and higher order anharmonic terms: the higher
order anharmonicity the smaller are corresponding non-central interactions as 
compared to central ones. Therefore, as a first step, only central forces may 
be accounted. 
One can show  (see Appendix) that the main effect of the non-central 
anharmonic interactions to the static local dynamics consists in 
renormalization of central elastic constants: $U^2V_4$ is replaced by 
$U^2(V_4 + 6V'_4)$. This allows one to improve the central-force  
approximation by replacing $V_4$ by $\bar{V}_4 = V_4 + 6V'_4$. 

We calculate a strong local vibration of light impurity or host 
atom (ion) situated at the origin of our reference frame. In this case 
solutions of classical equations of motion, 
corresponding to the local mode, satisfy the conditions: 
$|A_{0}| \gg |A_{\vec{n}}|$. This allows one to suppose that the mode is well 
localized on the atom at the site $n=0$. Then in the approximation of central 
forces 9 coordinates of atoms contribute to perturbation of the dynamical
matrix: 3 coordinates of the central atom and 6 
directed to this atom coordinates of the nearest neighbor atoms. 
We chose coordinates according to Fig. 4.
In this representation the impurity induced change of the
dynamical matrix $\bar{V}_2$ is 
\[ V = \left (\begin{array}{ccc} 
w_x&0&0\\\noalign{\medskip} 
0&w_y&0\\\noalign{\medskip} 
0&0&w_z\end{array} 
\right ) , \;\; 
w_{\alpha} = \left (\begin{array}{ccc} 
\beta_{\alpha}&-\gamma_{\alpha}&-\gamma_{\alpha}\\\noalign{\medskip} 
-\gamma_{\alpha}&\gamma_{\alpha}&0\\\noalign{\medskip} 
-\gamma_{\alpha}&0&\gamma_{\alpha}\end{array} 
\right ) . \] 
In harmonic approximation $\beta$ and $\gamma$ do 
not depend on $\alpha$: $\gamma = \Delta V_2$ is the change of the magnitude 
of the central elastic constants due to defect atom,  
$\beta = 2\gamma+\omega^2 (1-M/M_0)$, $\omega$ is the frequency of the normal 
mode, $M/M_0$ is the ratio of impurity and host atom masses. Quartic 
anharmonicity leads to amplitude dependent corrections of elastic constants 
and to their dependence of $\alpha$: 
\begin{equation} 
\beta_{\alpha} = \omega^2{\Big (}1- \frac{M}{M_0}{\Big )}+
2\gamma_{\alpha},\quad 
\gamma_{\alpha} = \Delta V_2+\frac{1}{2}V_4 A^2_{\alpha}, 
\end{equation} 
$A_{\alpha}$ is the $\alpha$'s Cartesian component of the amplitude of the 
local mode. 
Perturbed Green's functions are 
\[ G = \left (\begin{array}{ccc} 
G_x&0&0\\ \noalign{\medskip} 
0&G_y&0\\ \noalign{\medskip} 
0&0&G_z\end{array}
\right ) , \;\;
G_{\alpha} = \left (\begin{array}{ccc} 
G_{\alpha 11}&G_{\alpha 12}&G_{\alpha 12}\\ \noalign{\medskip} 
G_{\alpha 12}&G_{\alpha 22}&0\\ \noalign{\medskip} 
G_{\alpha 12}&0&G_{\alpha 22}\end{array} 
\right ) , \] 
where 
$$
G_{\alpha nn'} = ([I-G^{(0)}w_{\alpha}]^{-1}G^{(0)})_{nn'},
$$ 
\begin{equation} 
G^{(0)}_{nn} =  
\sum_{\nu \vec{k}}\frac{e_{n \nu \vec{k}}^2} 
{\omega^2-\omega^2_{\nu\vec{k}}}, \\\;\;\; 
G^{(0)}_{12} = 
\sum_{\nu \vec{k}}\frac{e_{1\nu \vec{k}} e_{2\nu \vec{k}}\cos{(k_{x}d)}} 
{\omega^2-\omega^2_{\nu\vec{k}}}, 
\end{equation} 
are Green's functions of perfect lattice, $d$ is the lattice constant, 
$\vec{k}$ is the wave vector of phonon, $\nu$ is the phonon branch, 
$\omega_{\nu\vec{k}}$ is the frequency of phonon, $e_{n \nu\vec{k}}$ is the 
projection of the polarization vector of phonon onto $x$-component of the 
atom. We use for the Green's functions $G^{(0)}$ their values 
calculated within shell model\cite{shell}. 

The frequency of the ALM, which depends on the amplitude of the mode, is given
by the position of the pole of the Greens function $G_{\alpha 00} (\omega)$
on the real axis of $\omega$.
Energy of the local mode depends on the amplitude via dependence of the
$\omega_l$ and directly:
\begin{equation} 
E_l \simeq \frac{1}{2}M\omega_l^2 A_l^2 + 
\frac{1}{64}\bar{V}_4(A_x^4+A_y^4+A_z^4) 
\end{equation} 
($A_l^2 = A_x^2+A_y^2+A_z^2$).

In a perfect lattice the perturbation matrix $\tilde{V}
\sim \bar{V}_4 A_{\alpha}^2$. This perturbation can lead to 
appearance of the ALMs. In three dimensional crystals they appear only 
if the local perturbation is strong enough. This means that
there is an minimal (critical) amplitude of the central atom $A_0$ for 
appearance of an ALM \cite{flach1}.
  
We performed calculation of the ALMs and ILMs associated with light 
$F^-$ and $Na^-$ ions in 
different alkali halide crystals. Results of some of our calculations are 
presented  on Figs. 3 and 4.
One sees that frequency of the ALM of light ion ${\rm F^-}$ in ${\rm KCl}$ 
rather strongly depends on amplitude of the mode and on the
crystallografic direction of the vibration. Thereat
the smaller is the distance to the nearest neighbor ($nn$) atom in direction 
of vibration the larger is effect of anharmonicity at the same amplitude. 
The reason for
this correlation is rather obvious: central anharmonic forces, which  
dominate in the anharmonic repulsive interactions, are stronger for 
smaller $nn$ distance. One also sees that the ILM on light host ion 
${\rm Na^+}$ in ${\rm NaI}$ appear already at rather 
moderate amplitude $\sim 0.4 \AA$. Thereat with increasing of amplitude
the mode first appear in (100) direction, then
in (110) direction and then in (111) direction. The reason 
for such directional dependence is given above: anharmonic 
interaction is stronger for smaller $nn$ distance.
Note that the decay time of the modes due to two-phonon emission, as it was
shown in \cite{quantmode} also strongly depends on the crystallografic 
direction.

In conclusion, the method of calculation of anharmonic local modes  both, 
ILMs (in pure crystals)  and ALMs (in doped crystals) is proposed. The 
method allows the reduction of the problem of nonlinear local dynamics 
to the linear inverse problem of phonons scattering on a 
local potential, caused by the local mode. 
New analytical description of the 
even and odd strongly localized ILMs in the monatomic chain is given.
Results of numerical calculations of anharmonic local modes of light ions
in pure and impure alkali halide crystals are presented.  
It is found that the ILM on a light host ion appears already at rather 
moderate amplitude (e.g $\sim 0.4 \AA$ in  ${\rm NaI}$).
Both, ILMs and ALMs in alkali-halide crystals strongly depend
on the crystallographic direction. 

\section{Acknowledgment} 
This research was supported by the Estonian Science Foundation, Grant 
No.~2274, DAAD Grant HSPS, and partially by TEMPUS PHARE Project 
S\_JEP-11212-96. V.H. thanks BTU, Cottbus for hospitality; 
D.N. thanks NORDITA for hospitality and financial support.

\subsection {Appendix: Non-central forces} 
Although non-central anharmonic interactions are remarkably weaker than 
central ones their number is larger. Therefore it is of interest to account 
them. These interactions switch-on into the perturbation $\tilde{V}_2$
all 21 Cartesian coordinates of the 
central atom and its 6 nearest neighbors. It is convenient 
to chose these coordinates as follows (see Fig.~1): 
${q}_{1} = {x}_{0}$, ${q}_{2} = {y}_{0}$, ${q}_{3} = {z}_{0}$, 
${q}_{4} = {x}_{1_{x}}$, ${q}_{5} = {y}_{1_{x}}$, ${q}_{6} = {z}_{1_{x}}$,  
${q}_{7} = {x}_{-1_{x}}$, ${q}_{8} = {y}_{-1_{x}}$, ${q}_{9} = {z}_{-1_{x}}$, 
${q}_{10} = {1}_{1_{y}}$, ${q}_{11} = {x}_{1_{x}}$, ${q}_{12} = {z}_{1_{y}}$, 
${q}_{13} = {y}_{-1_{y}}$, ${q}_{14} = {x}_{-1_{y}}$, 
${q}_{15} = {z}_{-1_{y}}$, ${q}_{16} = {z}_{1_{z}}$, ${q}_{17} = {x}_{1_{z}}$,
${q}_{18} = {y}_{1_{z}}$, ${q}_{19} = {z}_{-1_{z}}$,  
${q}_{20} = {x}_{-1_{z}}$, ${q}_{21} = {y}_{-1_{z}}$; 
$U = \sqrt{U_{x}^{2}+U_{y}^{2}+U_{z}^{2}}$. In this representation  
the perturbation matrix of lattice dynamics equals  
\[ V = \left (\begin{array}{cccc} 
\nu_0&-\bar{\nu}_1&-\bar{\nu}_2&-\bar{\nu}_3\\\noalign{\medskip} 
-\bar{\nu}_1^{\top}&\tilde{\nu}_1&0&0\\\noalign{\medskip} 
-\bar{\nu}_2^{\top}&0&\tilde{\nu}_{1'}&0\\\noalign{\medskip} 
-\bar{\nu}_3^{\top}&0&0&\tilde{\nu}_{1''}\end{array} 
\right ) , \] 
where $\nu_0 = \tilde{\beta} I_3$ is $3$ x $3$ matrix,   
$\bar{\nu}_{i^{(','')}} =(1,1)$ x $\nu_{i^{(','')}}$ are $3$ x $6$ matrixes, 
$\tilde{\nu}_1 = I_2$ x $\nu_1$ are $6$ x $6$ matrixes, $I_n$ is $n$x$n$-unit 
matrix, $\gamma$ and $\gamma'$ are changes of central and non-central elastic 
springs due to the impure central ion, 
$\tilde{\beta} = \omega^2 (1-M/M_0) + 2\gamma + 4\gamma'$,  
\[ \nu_1 = \left (\begin{array}{ccc} 
\gamma & 0 & 0 \\
\noalign{\medskip}
0 & \gamma' & 0 \\
\noalign{\medskip}
0 & 0 & \gamma'
\end {array}\right ),\;\; 
\nu_2 = \left (\begin{array}{ccc} 
0 & \gamma' & 0 \\
\noalign{\medskip}
\gamma & 0 & 0 \\
\noalign{\medskip}
0 & 0 & \gamma'
\end{array}\right ) ,\;\; 
\nu_3 = \left (\begin{array}{ccc} 
0 & \gamma' & 0 \\
\noalign{\medskip}
0 & 0 & \gamma' \\
\noalign{\medskip}
\gamma & 0 & 0 
\end{array}\right ) .\]
In harmonic approximation $\nu_1 = \nu_{1'} = \nu_{i''}$. Quartic 
anharmonicity causes amplitude-dependent renormalization of the elastic 
springs and leads to the following corrections of 
(additions to) given above matrixes $\nu$:   
\begin{enumerate}
\item (100)-direction: $\nu'_{1'} = \nu'_{1''} = \nu'_2$,   
\[ \nu'_0 = \left (\begin{array}{ccc} 
\zeta_1&0&0\\\noalign{\medskip}0&\zeta_2&0\\\noalign{\medskip}0 
&0&\zeta_2\end {array}\right ),\;\; 
\nu'_1 = \left (\begin{array}{ccc} 
\delta_1&0&0\\\noalign{\medskip}0&\delta_2&0\\\noalign{\medskip}0 
&0&\delta_2\end {array}\right ),\; 
\nu'_2 = \left (\begin{array}{ccc} 
0&0&0\\\noalign{\medskip}0&0&0\\\noalign{\medskip}0 
&0&\delta_2\end{array}\right ) ,\;\; 
\nu'_3 = \left (\begin{array}{ccc} 
0&0&0\\\noalign{\medskip}0&0&\delta_2\\\noalign{\medskip}0 
&0&0\end{array}\right ) ,\] 
where $\delta_1 = U^2(V_4 + 6V'_4)/4, 
\delta_2 = U^2V'_4/4, \zeta_1 = 2\delta_1, \zeta_2 = 4\delta_2$;  
energy correction equals $E'_l = U^4(V_4 + 6V'_4)/64$;   
\item (110)-direction: $\nu'_{1'} = \nu'_1$,  
\[ \nu'_0 = \left (\begin{array}{ccc} 
\xi_1&\bar{\sigma}&0\\\noalign{\medskip}\bar{\sigma} 
&\xi_1&0\\\noalign{\medskip}0 
&0&\xi_2\end {array}\right ),\; 
\nu'_1 = \left (\begin{array}{ccc} 
\sigma_1&\sigma_2&0\\\noalign{\medskip}\sigma_2&\sigma_2&0\\\noalign{\medskip}0
&0&\sigma_2\end {array}\right ),\; 
\nu'_2 = \left (\begin{array}{ccc} 
\sigma_2&\sigma_2&0\\\noalign{\medskip}0&0&0\\\noalign{\medskip}\sigma_1 
&\sigma_2&\sigma_2\end{array}\right ),\; 
\nu'_3 = \left (\begin{array}{ccc} 
0&0&0\\\noalign{\medskip}\sigma_2&0&0\\\noalign{\medskip}0 
&0&0\end{array}\right ),\; 
\nu'_{1''} = \left (\begin{array}{ccc} 
\sigma_2&0&0\\\noalign{\medskip}0&0&0\\\noalign{\medskip}0 
&0&0\end{array}\right ),\]  
where$\sigma_1 = 
U^2(V_4 + 7V'_4)/8, 
\sigma_2 = U^2V'_4/8,  \bar{\sigma} = 4\sigma_2,  
\xi_1 = 2\sigma_1 + 6\sigma_2, \xi_2 = 6\sigma_2$;  
energy correction equals $E'_l = U^4(V_4 + 12V'_4)/128$;   
\item (111)-direction: $\nu'_{1'} = \nu'_{1''} = \nu'_1$,  
\[ \nu'_0 = \left ( \begin{array}{ccc} 
\phi & f & f \\ 
\noalign{\medskip} f & \phi & f \\
\noalign{\medskip} f & f & \phi
\end{array}\right ),\;\; 
\nu'_1 = \left ( \begin{array}{ccc} 
\kappa_1 & \kappa_2 & \kappa_2 \\
\noalign{\medskip} \kappa_2 & \kappa_3 & 0 \\
\noalign{\medskip} \kappa_2 & 0 &\kappa_3
\end{array}\right ),\;\; 
\nu'_2 = \left ( \begin{array}{ccc} 
\kappa_2 & \kappa_3 & 0 \\ 
\noalign{\medskip}\kappa_1&\kappa_2&\kappa_2 \\
\noalign{\medskip}\kappa_2 & 0 &\kappa_3
\end{array}\right ) ,\;\; 
\nu'_3 = \left( \begin{array}{ccc} 
\kappa_2 & \kappa_3 & 0\\ 
\noalign{\medskip}\kappa_2 & \kappa_3 & \delta_2 \\
\noalign{\medskip}\kappa_1 &\kappa_2&\kappa_2
\end{array}\right ) ,\] 
where $\kappa_1 = U^2(V_4 + 8V'_4)/12, 
\kappa_2 = U^2V'_4/6, \kappa_3 = \kappa_2/2, \phi = 2\kappa_1 + 4\kappa_2, 
f = 8\kappa_3$; energy correction equals $E'_l = U^4(V_4 + 18V'_4)/192$.
\end{enumerate}
Corrections $\sim V''_4$ are neglected; they are at least one order of 
magnitude smaller than smallest accounted correction $\sim V'_4$.   
As one sees the main effect of the non-central anharmonic interactions to the
local dynamics consists in renormalization of central elastic constants:
$A^2V_4$ is replaced by $A^2(V_4+ 6V'_4)$.

\newpage

\begin{center}
{\Large {\bf Figure captions}}
\end{center}

\bigskip

\noindent
{\bf Figure 1.} $\beta$ (solid lines) and $A_l$ (dot-dashed lines) vs. 
$\omega_l$
for even and odd modes. We used for calculations $\omega_M=1$ and
$V^{(4)}/M_0=1$.

\noindent
{\bf Figure 2.} Atomic displacements, which contribute to the 
perturbation of  the dynamical matrix: 
a) account of central forces --  9 displacements;
b) account of central and noncentral forces -- 21 displacements.

\noindent
{\bf Figure 3.} Frequency of the localized mode ($w_l/w_m$) vs.
amplitude (in $\AA$). $KCl:F$, [100] -- solid line, [110] -- dashed line,
[111] -- dot-dashed line.

\noindent
{\bf Figure 4.}  Frequency of the localized mode ($w_l/w_m$) vs.
amplitude (in $\AA$). $NaI$, [100] -- solid line, [110] -- dashed line,
[111] -- dot-dashed line.


\newpage
\begin{thebibliography}{99} 
\bibitem{ovchi}  
A.A.Ovchinnikov, Zh. Eksp. Teor. Fiz. {\bf 57}, 263 (1969) 
[Sov. Phys. JETP {\bf 30} 147 (1970)]; 
A.A.Ovchinnikov and N.S.Erihman, Usp. Fiz. Nauk {\bf 138}, 290 (1982) 
[Sov. Phys. Usp. {\bf 25} 738 (1982)]. 
\bibitem{kosevich} 
A.M.Kosevich and A.S.Kovalev, Zh. Eksp. Teor. Fiz. {\bf 67}, 1793 (1974) 
[Sov. Phys. JETP {\bf 40}, 891 (1974)]. 
%
\bibitem{dolgov}  
A.S.Dolgov, Fiz. Tverd. Tela (Leningrad) {\bf 28}, 1641 (1986) 
[Sov. Phys. Solid tate {\bf 28}, 907 (1986)]. 
\bibitem{sivtak}  
A.J.Sievers, S.Takeno, Phys. Rev. Lett. {\bf 61}, 970 (1988). 
\bibitem{page}  
J.B.Page, Phys. Rev. B {\bf 41}, 7835 (1990).   
\bibitem{zavt}  
G.S.Zavt {\it et al.}, Phys. Rev. E {\bf 47}, 4108 (1993). 
\bibitem{flach}
S.Flach, C.R.Willis, Physics Reports {\bf 295},181 (1998).
\bibitem{hizhrev}  
V.Hizhnyakov, Phys.Rev. B {\bf 53}, 13981 (1996); 
{\it Proceedings of the  XIIth Symposium on the Jahn-Teller  
Effect} [Proc. Estonian Ucad. Sci. Phys. Math. {\bf 44}, 364, (1995)].    
\bibitem{hizhnev}  
V.Hizhnyakov, D.Nevedrov, Proc. Estonian Ucad. Sci. Phys. Math. {\bf 44},  
376 (1995); 
V.Hizhnyakov and D.Nevedrov, Z. Phys. Chem. {\bf 201}, 301 (1997); 
V.Hizhnyakov and D.Nevedrov, Pure and Appl. Chem. {\bf 69}, 1195 (1997). 
\bibitem{zhizh} 
V.Hizhnyakov, Z. Phys. B, {\bf 104} 675 (1997).
\bibitem{chain} 
V.Hizhnyakov and D.Nevedrov, Phys. Rev. B {\bf 56}, R2809 (1997). 
\bibitem{euro}
V.Hizhnyakov, Europhys. Lett. {\bf 45} (4), 508 (1999).
\bibitem{bilz}
H.Bilz, W.Kress, {\it Phonon Dispersion Relations in
Insulators} (Springer, Berlin, 1979).
\bibitem{shell} 
A.D.B.Woods, W.Cochran, and B.N.Brockhouse,
Phys. Rev. {\bf 119}, 980 (1960);
A.D.Woods, B.N.Brockhouse, R.A.Cowley, and W.Cochran,
Phys. Rev {\bf 131}, 1025 (1963);
R.A.Cowley, W.Cochran, B.N.Brockhouse, and A.D.B.Woods,
Phys. Rev {\bf 131}, 1030 (1963).
\bibitem{maradu}  
A.A.Maradudin {\it et al.} {\it Theory of Lattice Dynamics in  
Harmonic Approximation} (Academic, New York, 1963);   
A.A.Maradudin, {\it Theoretical and 
Experimental Uspects of the Effects of Point Defects and Disorder of the  
Vibrations of Crystals} (Academic, New York 1966).  
\bibitem{kristofel}
N.N.Kristofel, {\it Theoriya primesnyh centrov ,alyh radiusov
v ionnyh kristallah}  (Nauka, Moscow, 1974).
\bibitem{economou} 
E.N.Economou, {\it Green's Functions in Quantum Physics} 
(Springer-Verlag, Berlin, 1983). 
\bibitem{quantmode}
V.Hizhnyakov and D.Nevedrov, Z. Phys. Chem. {\bf 201}, 301 (1997);
Pure and Appl. Chem. {\bf 69}, 1195 (1997); Physica B {\bf 263-264} 762 (1999).

\end{thebibliography}
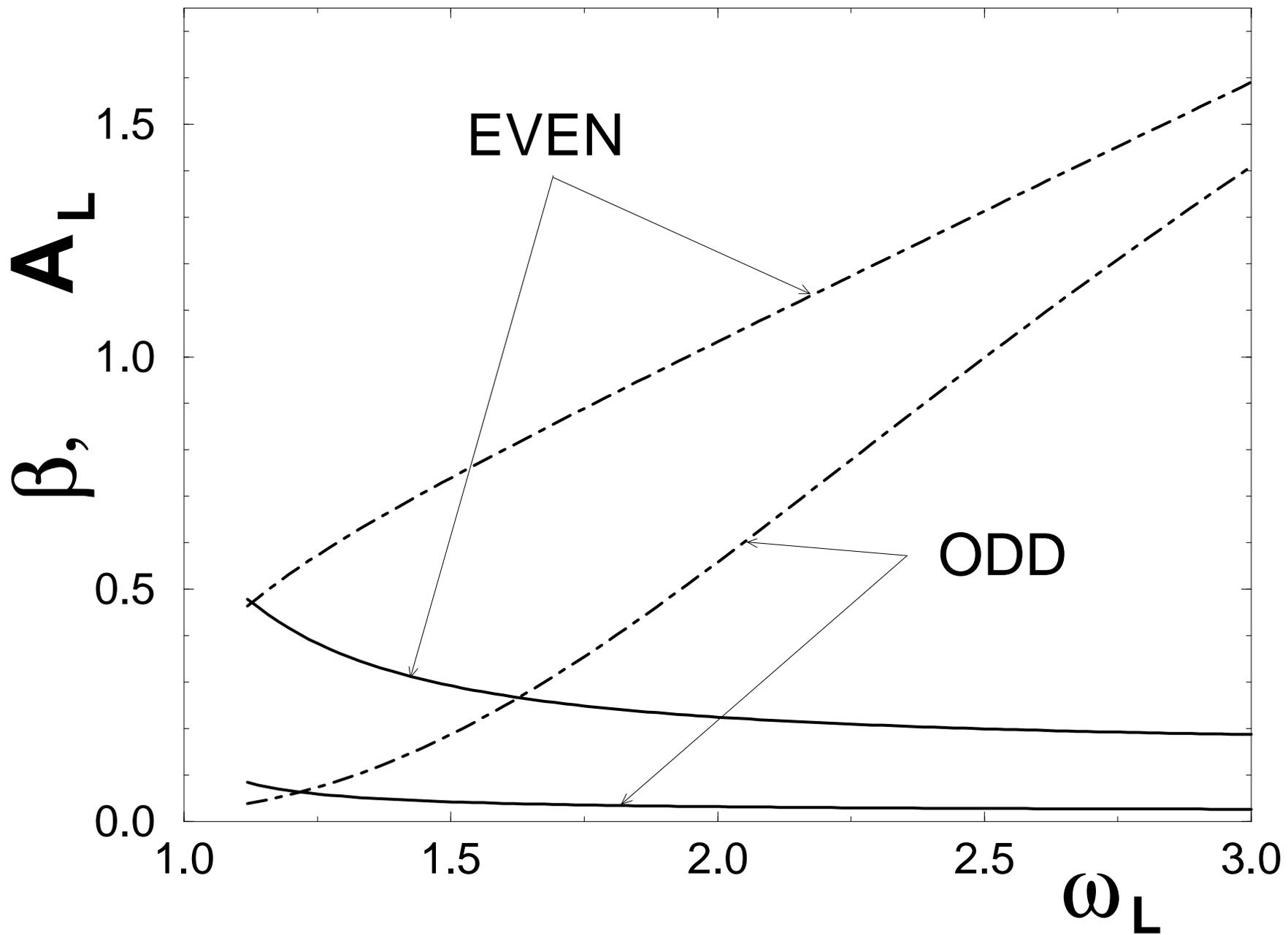
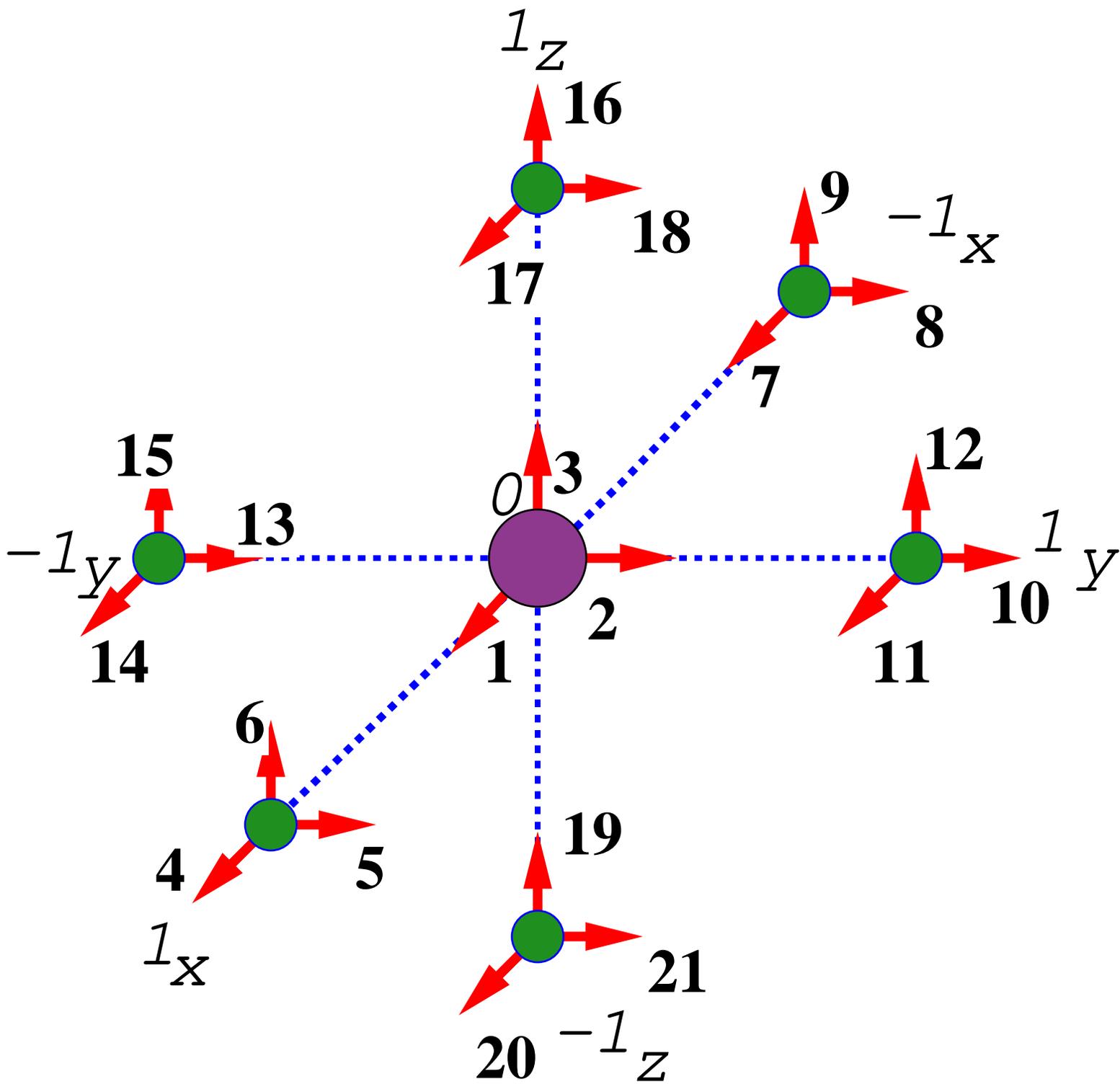
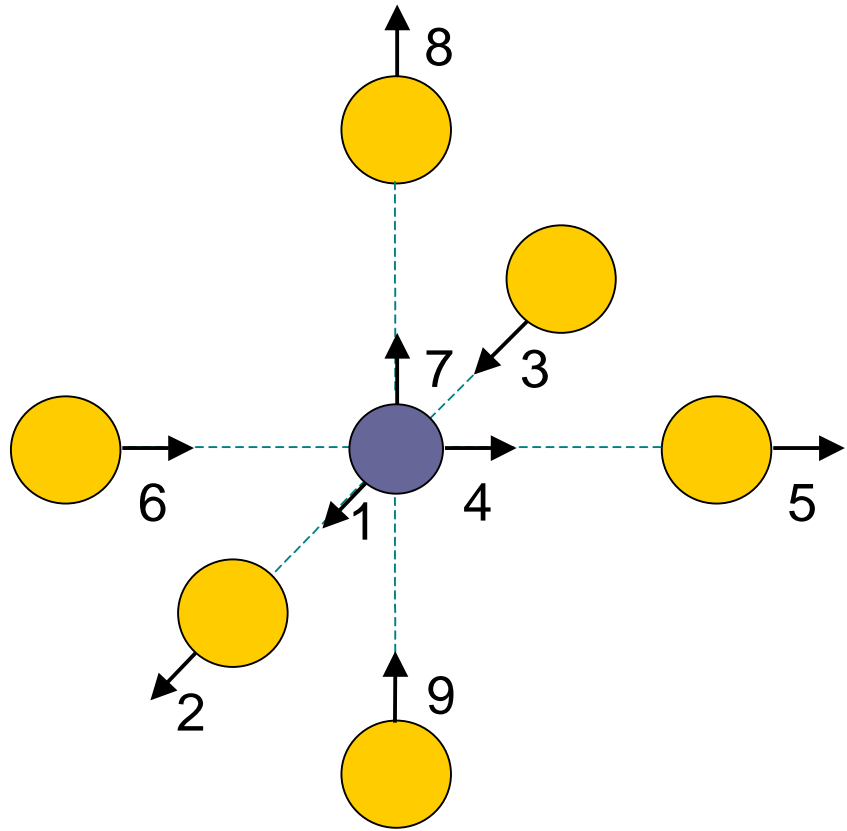
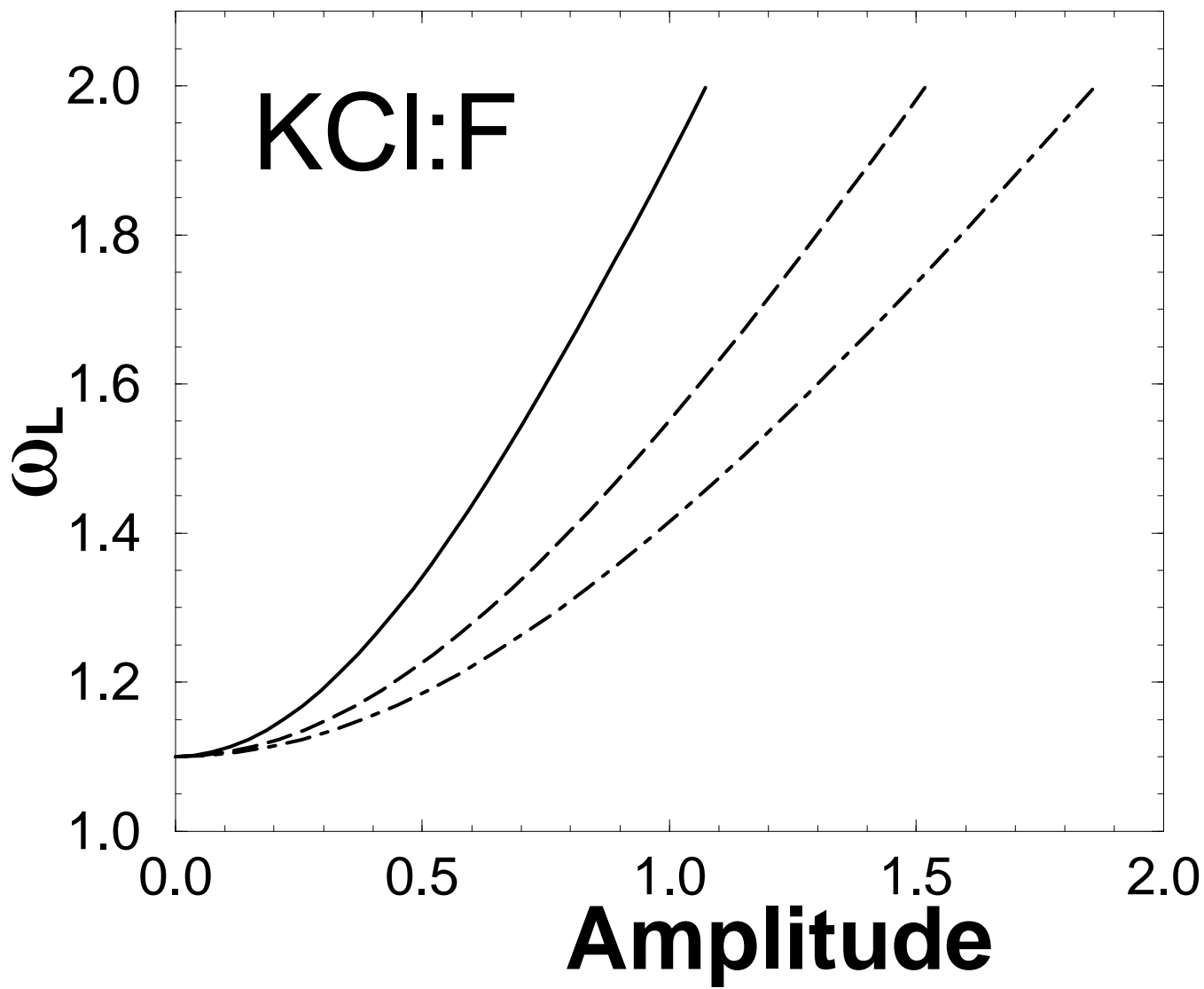
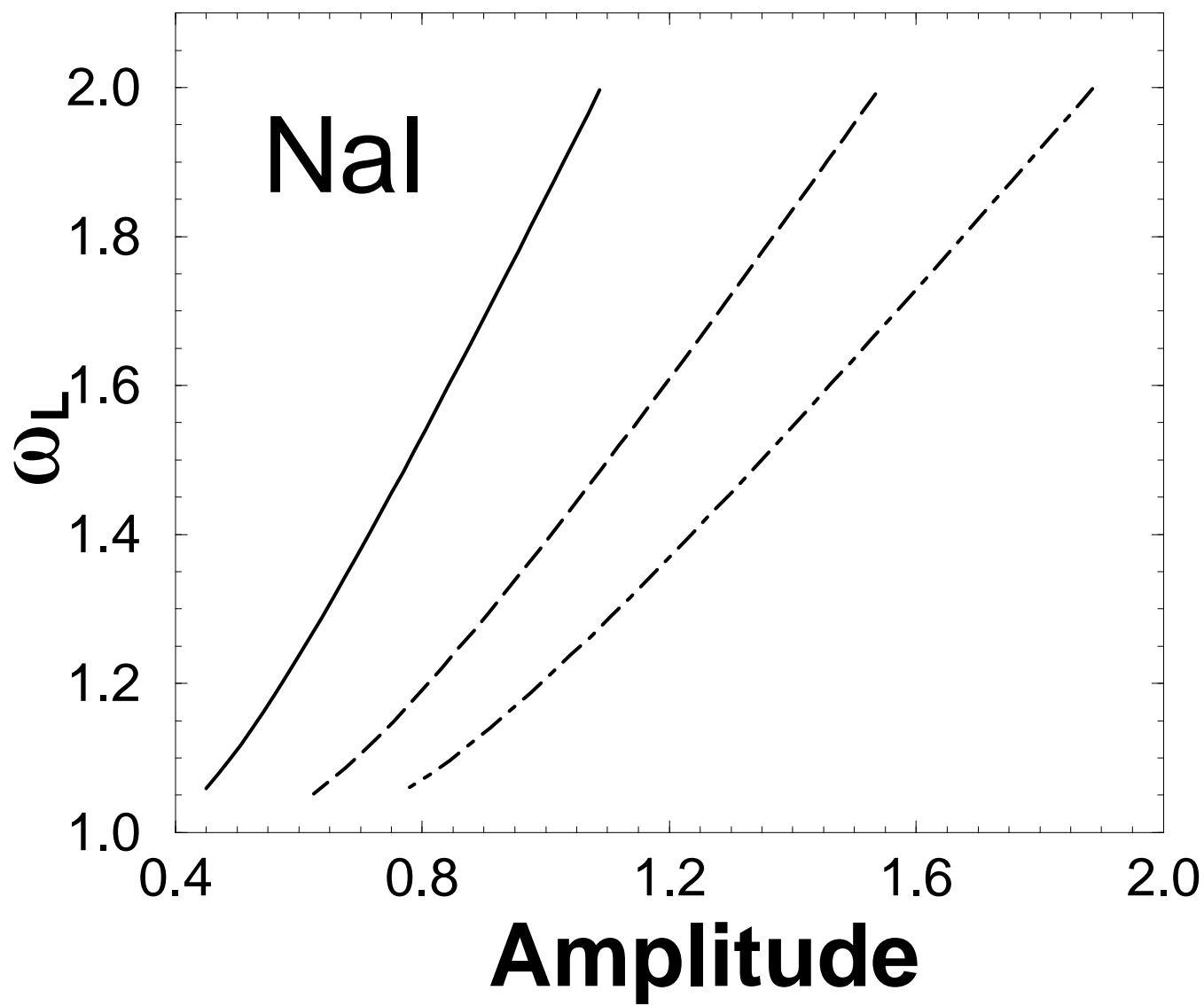
\end{document}